\documentclass[prl,showpacs,noshowkeys,amsmath,superscriptaddress,twocolumn,letterpaper]{revtex4}
\usepackage{ae,aecompl}
\usepackage{longtable}
\usepackage{graphicx}
\newcommand{\be}{\begin{eqnarray}}
\newcommand{\ee}{\end{eqnarray}}

\newcommand{\eff}{{e\!f\!\!f}}
\begin{document}

\title{Kramers Theory for Conformational Transitions of Macromolecules} 
\author{Marcello Sega}
\affiliation{Frankfurt Institute for Advanced Studies, Ruth-Moufang-Str.~1, D-60435 Frankfurt, Germany}
\email{sega@fias.uni-frankfurt.de}
\author{Pietro Faccioli}
\affiliation{Dipartmento di Fisica, Universit\`a di Trento, Via   Sommarive 14, I-38050 Povo (Trento), Italy}
\affiliation{I.N.F.N Gruppo Collegato di Trento}
\author{Francesco Pederiva}
\affiliation{Dipartmento di Fisica, Universit\`a di Trento, Via   Sommarive 14, I-38050 Povo (Trento), Italy}
\affiliation{I.N.F.M-Democritos, National Simulation Center, Trieste, Italy}
\author{Henri Orland}
\affiliation{Service de Physique Th\'eorique, Centre d'Etudes de Saclay, F-91191 Gif-sur-Yvette Cedex, France}

\date{\today}
\begin{abstract}
We consider the application of  Kramers theory to  the microscopic calculation of rates of conformational transitions of macromolecules. The main difficulty in such an approach is  to locate the transition state in a huge configuration space. We present a method which identifies the transition state along the most probable reaction pathway. 
It is then possible to microscopically compute the activation energy, the damping coefficient, the eigenfrequencies at the transition state  and obtain the rate, without any {\em a priori} choice of a reaction coordinate. 
Our theoretical results are tested against the results of Molecular Dynamics simulations for transitions in a 2-dimensional double well and for the cis-trans isomerization of a linear molecule.
\end{abstract}
\maketitle
\noindent 
The kinetics of conformational changes of macromolecules is believed  to provide important information about the underlying mechanisms involved in such reactions.
In such a context, rates are the fundamental observables. Not only they provide direct tests for theoretical calculations, but they also encode information about the structure of the important reaction pathways. For example, by $\phi$-value analysis it is possible to identify the residues which are structured at the transition state~\cite{phi}. 

Kramers theory and its multidimensional generalization offer a scheme to compute the  transition rates for bistable molecular systems.
In such a formalism  \cite{Kramers:1940,Hanggi:1990,Landauer:1961}, 
the transition rate of a particle in an external potential $U(x)$ in $N$ dimensions, from the meta-stable state $b$ across the saddle-point $o$ can be written, in the strong friction regime, as 
\be
k_K =\frac{\omega_s}{ \gamma \tau_{eq}}, 
\quad \tau_{eq}^{-1} \equiv \frac{1}{(2\pi)}\,\frac{\prod_i^{N} \omega^b_i}{ \prod_j^{N-1}
{\omega^o_j}}\, e^{- \frac{[U(o)-U(b) ]}{k_BT}},
\label{Kramers}
\ee
where  $\gamma$ is the
friction coefficient, $\omega_s$ is the angular frequency of the
single unstable mode at the saddle-point and $\omega^b_i$  and $\omega^o_j$ are the stable frequencies in $b$ and in $o$, respectively. The ratio $\omega_s / \gamma$  is often called the (adimensional) damping factor. It is responsible for lowering the actual rate from the theoretical upper limit, $\tau_{eq}^{-1}$, given by transition state theory.

Kramers theory has been successfully applied to more complicated chemical reactions involving macromolecules \cite{Hanggi:1990}, such as two-state protein folding.
In this context, Eq.(\ref{Kramers}) is used as a phenomenological prescription in which $x$ is a set of reaction coordinates,  $U(x)$ is the corresponding potential of mean force (free energy) and $\gamma$ is the effective friction at the transition state. 

In the present work we adopt a different strategy: The configuration $x$ specifies the microscopic degrees of freedom of the molecule (e.g. the atom or residue coordinates) and $U(x)$ is the interaction potential with implicit solvent \cite{Berne:1998}.
The main difficulty in such an approach to chemical reaction rates, is that it requires to know the location of the saddle-point state $x=o$ in a large configuration space. This information is needed to determine the activation energy, the damping factor and the eigenfrequencies, which are in turn needed to estimate the rate constant.
In practice, the identification of the transition state in molecular reactions represents a very challenging task and Kramers formula cannot be directly applied.   

The key point of this work is to show that the problem of finding the transition state in  two-state conformational reactions of  macro-molecules  can be efficiently solved using the recently developed Dominant Reaction Pathways formalism \cite{Faccioli:2006, Sega:2007, Autieri:2008,Faccioli:2008}.
This formulation  of the stochastic dynamics leads to an impressive computational simplification of the problem of finding the most important reaction pathways in high dimensional systems \cite{Faccioli:2006}. The reason is that the dominant reaction pathway is sampled at equally-spaced displacement steps, rather than using constant time steps.  In thermally activated reactions, due to the decoupling of time scales, the difference between these two samplings is huge. In particular, for the folding of a  polypeptide chain, the number of displacement discretization steps is of order $30-100$ \cite{Faccioli:2006,Sega:2007}. This number should be compared with the order  $10^{12}$ steps which would be required to describe the same reaction using constant time steps. 
In our recent work ~\cite{Faccioli:2006,Sega:2007,Faccioli:2008}, we have shown that it is possible to determine the most statistically important reaction pathways in conformational transitions of amino-acid chains.
In this Letter, we show how to perform the atomistic calculation of Kramers  reaction rates in macromolecular transitions by using the most probable paths, with available computers.

We begin by briefly reviewing the Dominant Reaction Pathways approach (for a detailed and self-contained introduction,  see \cite{Autieri:2008}). For the sake of simplicity, we present all formulas in the case of the diffusion in a one-dimensional external potential.  The  generalization to the multidimensional case is straightforward. 
Let us therefore consider the overdamped Langevin dynamics of a system with coordinate $x$ in a potential~$U(x)$:
\begin{equation}
\frac{\partial x}{\partial t} = - \frac{D}{k_B T} \nabla U(x) + \eta(t),
\label{lan}
\end{equation}
where $D= k_B T/\gamma$ is the diffusion coefficient,  and $k_B T$ is the thermal energy.
$\eta(t)$ is a Gaussian noise with zero average and correlation given by 
$\langle\eta(t)\eta(t')\rangle= 2D\delta(t-t')$. 
Note that in the original Langevin equation an inertial term,
$m \ddot{x}$, appears.
However, in the case of the dynamics of biopolymers in water, the effect of such a term can be neglected for time scales larger than a fraction of picosecond \cite{orland}.
As is well known, the stochastic differential equation (\ref{lan}) generates a probability distribution $P(x,t)$ which obeys the Fokker-Plank Equation (FPE)
\be
\frac{  \partial P(x,t)}{\partial\,t}=D \nabla 
\left(\frac{\nabla  U(x)}{k_B T}P(x,t)\right)
+D\nabla^2P(x,t).\nonumber\\
\label{FPE}
\ee
In the study of noise-driven reactions, we are interested in transitions between two meta-stable states $a$ and $b$.  
The starting point of the Dominant Reaction Pathways approach is  to express the solution of the FPE, subject to the boundary conditions $x(0)=b$ and $x(t)=a$ in terms of a path integral:
\be
\label{path}
P(b,0;a,t)=e^{-\frac{U(b)-U(a)}{2 k_B T}}
\int_{b}^{a} \mathcal{D}x\, e^{-{S_{\eff}[x]}},
\label{PI}
\ee
where $S_{\eff}[x]=\int d\tau \left( \dot x^2/4D +V_{eff}[x]\right) $ is an effective action and the effective potential reads
$
V_{\eff}(x)=D/( 2 k_B T)^2 \left[( \nabla U(x))^2
-2 k_B T\nabla^2 U(x)\right] .
$ 

The most probable paths in configuration space contributing to (\ref{PI}) are called the {\it dominant reaction pathways} (DRP). They are those for which the exponential weight  $e^{-S_{eff}}$ is maximum, hence for which the effective action $S_{eff}$ is minimum. 
In our recent works ~\cite{Faccioli:2006,Sega:2007, Autieri:2008}  we have shown that the DRP  can be rigorously obtained by minimizing the effective Hamilton-Jacobi (HJ) functional
\be
S_{HJ}([x];b,a)\equiv\int_{b}^{a} d\ell \sqrt{D^{-1}\left(V_{\eff}[x(\ell)]-V_{\eff}(a)\right)}, 
\label{HJ}
\ee 
where $d\ell=\sqrt{\sum_i dx^2 }$ is a measure  of the elementary displacement along the reaction path. We note that the HJ functional does not depend on the diffusion coefficient $D$. Hence, the DRPs are not sensitive to the choice of the friction coefficient. This could be seen already at the level of the FPE (\ref{FPE}),  in which the choice of the diffusion constant  just sets the time scale.    

The transition state can be identified as the configuration for which the potential energy is maximum along the DRP, and is a saddle point on the full potential energy landscape.
Once the transition state $x_{ts}$ has been found, one can easily compute the eigenfrequencies and damping factor required to compute the rate from eq. (\ref{Kramers}). 
With this knowledge, it is possible to estimate the friction coefficient in the transition state, which enters in (\ref{Kramers}), by means of short MD simulations with explicit solvent.

\begin{figure}
\centering
\includegraphics[width=.65\columnwidth]{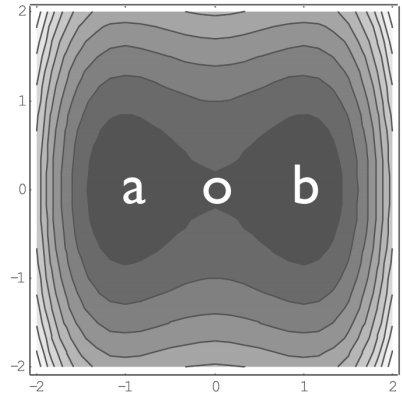}
\caption{Contour plot of the two-dimensional energy landscape used to test the method.
\label{2Dpl}
}
\end{figure}

Before showing examples of applications of the present method to specific systems, we first show
that the DRP contains much more information about the kinetics of the reaction than encoded in the Kramers rate formula. 
In fact, it is possible to use this method  to obtain microscopic parameter-free predictions  not only for the rate, but also for  the probability of performing an arbitrary number $q$ of transitions between $b$ and $a$, in a given time interval $t$.
We start from the work of Caroli {\it et al.}~\cite{Caroli:1981}, who  studied the diffusion in an asymmetric one-dimensional double-well, by  estimating the path integral  (\ref{PI})
in the so-called dilute instanton gas approximation. 
They obtained the expression for the transition probability  form $b$ to $a$ in a time $t$, with unconstrained  number of barrier crossings:
\be 
P(b,0;a,t) = \sqrt{\frac{U''(b)}{2\pi k_BT}} \frac{k_a}{k_b+k_a} \left\{1-e^{-(k_b +k_a)t}\right\},
\label{eq:totProb}
\ee
where 
\be
k_i=D\sqrt{\left|U''(o)\right|U''(i)} (2\pi k_BT)^{-1}e^{-\frac{U(o)-U(i)}{k_BT}}.
\label{ki}
\ee
Note that the relaxation to the equilibrium distribution is controlled by the rate
$(k_a+k_b)$, which is precisely Kramers rate. 
A generalization of this result to a fixed number $q$ of barrier crossing in time $t$ is made possible by exploiting the peculiar form of the path-integral solution, which allows to find the analytical solution for a generic double well. Here we show only the result for the symmetric double-well, which takes the particularly simple form
$
P^{(q)}(t) \propto {\left(k_b t\right)^{q} e^{-k_bt}}/ { q\,!}.
$
It is easy to check that after summing over all possible numbers of crossing, one recovers the correct unconditioned probability (\ref{eq:totProb})  with Kramers time
$\tau_k=(2k_b)^{-1}$ now appearing at the exponent.
These results can be generalized to bistable systems in higher dimensions, by replacing the expression in Eq.(\ref{ki}) for $k_b$ with 
$k_b = \omega_s/\gamma\tau_{eq} $, where  $\omega_s$ and $\tau_{eq}$ are defined as in Eq.(\ref{Kramers}).
\begin{figure}
\centering
\includegraphics[width=0.9\columnwidth]{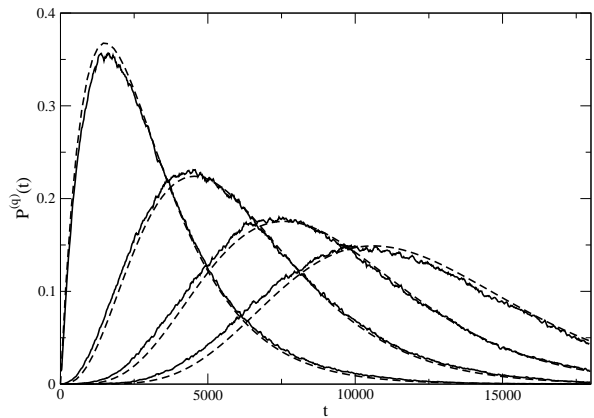}
\caption{
Comparison between Dominant Reaction Pathway prediction (dashed line) and MD simulations (solid line) for the $P^{(q})(t)$, with $q=1,3,5,7$,  in the two-dimensional double well of Fig.~\ref{2Dpl}. 
\label{fig:2d}
}
\end{figure}

We now illustrate how the present method is implemented in practice and we compare it with Molecular Dynamics (MD) simulations results, focusing on the generalized transition probabilities $P^{(q)}(t)$, since  they provide a very stringent test for the approach we have developed.

Consider the simple case 
of the  2-dimensional bistable  potential $U(x,y)/k_BT = A(x^2-1)^2 + B\, y^2$, plotted in Fig.~\ref{2Dpl} with $A=4$ and $B=15$.
In Fig.~\ref{fig:2d} we report the sampled histograms of $P^{(q)}(t)$
up to $q=7$, obtained from several MD simulations, and we  compare it with the corresponding theoretical curves obtained by the present analysis. 
In this case the  dominant trajectory is unique and  coincides with the straight line connecting the minima of $U(x,y)$. 
 
On the other hand, the construction of the histograms requires a prescription to record crossing events. We decided to set two thresholds, one on
the left  and one on the right of the barrier top, at
$x=\pm~0.4$, respectively. A crossing event is marked only when
the two thresholds are crossed in sequence. The histogram of the
transitions is updated each time the system is closer to $a$ than a given distance $\delta$.
The domain width $\delta$ and the integration
time-step have been set respectively to $0.02$ and $3\times10^{-4}$.

The  agreement between the analytic calculation based on the DRP and
the histograms obtained from the  MD simulation is excellent and holds for all the conditional probabilities, $P^{(1)}(t),...,P^{(7)}(t)$. We emphasize again the fact that the theoretical
curves are parameter-free and therefore the agreement represents a
compelling evidence, supporting the validity of the method. 

Let us now consider the application of the Dominant Reaction Pathways method to the kinetics of conformational transitions of a molecular system. It is important to stress the fact that  for sufficiently complex molecules, the position of the transition state in a conformational transition cannot in general be inferred {\it a priori}. This limitation prevents the direct application of Kramers theory. 
The aim of the present analysis is to check if the present approach  is able to find the correct saddle-point, and predict the $P^{(q)}(t)$ (and therefore Kramers rate).  
To this end, we consider the cis-trans isomerization of a toy molecule, 
in which the exact location of the transition state is known by construction.  
In particular, we used a model for a linear molecule composed of 8 interaction 
centers, with fixed bond lengths of 1~\AA, masses of 10 a.m.u., and torsional
potentials acting on the 5 dihedral angles $\phi_d $, of the form $U= C_d \left[1+\cos(n_d\phi_d-\phi_d^0)\right]$. The interaction parameters are $C_3=3kJ/mol$, $n_3=2$ and $\phi_3^0=90^\circ$ for the central dihedral $\phi_3$ and $C_d=10 kJ/mol$, $n_d=1$ and $\phi_d^0=0$ for the remaining four dihedral angles. The plane angles between every three consecutive atoms are kept fixed at $90^\circ$. This potential
has two minima, located at the \emph{cis} and \emph{trans} conformation relative to the central dihedral,
respectively. 

We computed the DRP by minimizing numerically the HJ action by means of a simulated annealing algorithm, starting from several  randomly generated arbitrary paths connecting the two states.  We then looked for the maximum of the potential energy along the reaction path to identify $x_{ts}$. The correct location of the saddle-point took only few minutes of CPU time and gave the same result, regardless of the random starting path used.

The next step consists in computing the equilibrium constant $1/\tau_{eq}$ which enters the 
expression of the $P^{(q})(t)$ and of the Kramers rate formula. If the only active degrees of freedom are the torsional angles (bond length and angles are kept fixed),  the Jacobian of the coordinate transformation
from Cartesian to internal ones does not depend on the internal
coordinates themselves~\cite{Pitzer:1946}. This fact can be used to
simplify the expression for the equilibrium rate constant for a
generic molecule\cite{Hanggi:1990} to the following:
\be
\frac{1}{\tau_{eq}} = \frac{1}{2\pi} \frac{|M^{-1/2} \nabla_x \ell(o)|}{|\nabla_\phi \ell(o)|} 
\sqrt{\frac{\prod_i^N \lambda^b_i}{\prod_j^{N-1} \lambda^o_j}} e^{-\frac{U(o)-U(b)}{k_BT}}.
\label{eq:internal}
\ee
Here $N$ is the number of the internal degrees of freedom, $M$ is the diagonal
matrix of the atom masses, and $x$ and $\phi$ indicate the Cartesian and internal
coordinates (dihedral angles in the present case), respectively. The (positive) eigenvalues of the Hessian matrix
$H_{ij}=\frac{\partial^2 U(\phi)}{\partial \phi_i \partial \phi_j}$ evaluated at the 
saddle point and in the starting well are 
denoted $\lambda^{o}$ and $\lambda^{b}$, respectively. 
The damping factor $\Gamma$  can be estimated\cite{Berne:1998} as
$
\Gamma = \frac{1}{k_B T}\frac{\Lambda}{\sqrt{\lambda^o_-}}
\frac{|\nabla_\phi \ell(o)|}{|M^{-1/2} \nabla_x \ell(o)|},
$
where $\Lambda$ is the negative eigenvalue of the $\hat{\mathbf{D}}(o)\mathbf{H}(o)$ matrix, $\hat{\mathbf{D}}(o)$ is the diffusion tensor in internal coordinates and, $\lambda^o_-$ is the negative eigenvalue of the Hessian matrix at the transition state. The Kramers rate has to be estimated in the moderate-strong friction regime, where the adimensional damping factor $\Gamma\simeq 0.5$. The computation of  $k_K = \left[\sqrt{\left(2\Gamma\right)^{-2} +1} - \left(2\Gamma\right)^{-1}\right] \tau_{eq}^{-1}$, resulted in a rate of $0.034$ ps$^{-1}$.

\begin{figure}
\centering
\includegraphics[width=0.9\columnwidth]{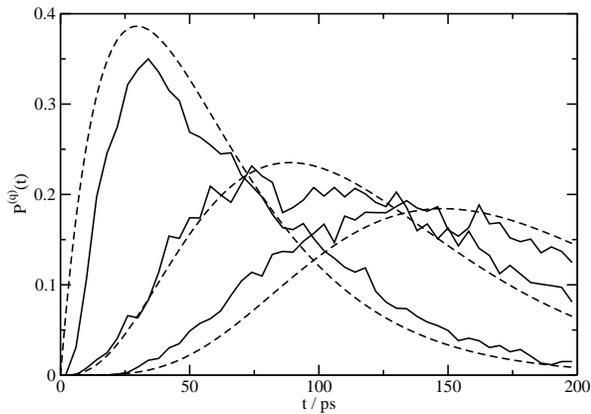}
\caption{
Comparison between the $P^{q}(t)$ functions for the cis-trans isomerization of the toy molecule, obtained from MD simulations (solid lines) and from the DRP  (dashed lines).}
\label{fig:chair}
\end{figure}

As before, we are interested in assessing the reliability of this approach, by comparing the prediction of the generalized transition probabilities with the results of MD simulations.  We integrated the full Langevin equation using a diffusion coefficient of $0.1$\AA$^2~$ps$^{-1}$ and 
integration time step of $0.001$~ps, at a temperature
of $300~K$. The histograms obtained from $10^{10}$ integration time-step are shown in Fig.~\ref{fig:chair},
along with the prediction obtained from the present method. The thresholds and domain width $\delta$ for the cosine of the torsional angles have been chosen to be $\pm0.5$ and 0.05, respectively.
The agreement between the theoretical predictions and the results of numerical simulations is quite good, given the approximations involved in the estimate of the rate constant.
The calculation of the $P^{(q)}(t)$  starting from the DRP  took only minutes of CPU time. On the other hand, approximatively 150 CPU hours were required to reconstruct the same curves from histograms obtained from MD simulations. Note that such a gain was observed also for reactions in more realistic systems, using empiric atomistic force fields~\cite{Sega:2007}. 

In conclusion, in this work we have developed a parameter-free method
to compute Kramers rate at the microscopic level, in high-dimensional
systems exhibiting two-state kinetics.
Our  theoretical expressions for the generalized transition
probabilities $P^{(q)}(t)$ have been found in excellent agreement with the results
of MD simulations performed in a two-dimensional double well and in the cis-trans isomerization of a simple molecule. 
This approach is very accurate and leads to a huge computational gain with respect to MD simulations, and makes  it possible to perform calculations of rates for systems in which MD techniques are not presently feasible.

Computations were performed on the HPC facility at the Department
of Physics of the Trento University and at the Frankfurt Center for
Scientific Computing, whose support we gratefully acknowledge.
We thank W. Eaton and A.Szabo for important discussions.

\end{document}